\documentclass{article}
\usepackage{spconf}

\usepackage{amsmath, bbm}
\usepackage{amssymb}
\usepackage{amsfonts}
\usepackage{amsthm}
\usepackage{algorithm}
\usepackage{algorithmic}
\usepackage{nicefrac}
\usepackage{bm}
\usepackage{hyperref}
\usepackage{cite}
\usepackage[utf8]{inputenc} 
\usepackage[T1]{fontenc}
\usepackage{graphicx}
\usepackage[font=small]{caption}
\usepackage{tikz}
\usetikzlibrary{positioning}
\usepackage{pgfplots}
\usepackage{epstopdf}
\usepackage[labelformat=simple]{subcaption}
\usepackage{multirow}
\usepackage{lipsum}
\usepackage{mathtools}

\long\def\comment#1{}

\input{mysymbol.sty}

\title{Joint Task Offloading and Routing in Wireless Multi-hop Networks Using Biased Backpressure Algorithm}

\name{\parbox{\textwidth}{\centering Zhongyuan Zhao$^\star$, Jake Perazzone$^\ddag$, Gunjan Verma$^\ddag$, Kevin Chan$^\ddag$, \\
\textit{Ananthram Swami}$^\ddag$, and \textit{Santiago Segarra}$^\star$}
\thanks{
    Research was sponsored by the DEVCOM ARL Army Research Office and was accomplished under Cooperative Agreement Number W911NF-24-2-0008. 
    The views and conclusions contained in this document are those of the authors and should not be interpreted as representing the official policies, either expressed or implied, of the DEVCOM ARL Army Research Office or the U.S. Government. 
    The U.S. Government is authorized to reproduce and distribute reprints for Government purposes notwithstanding any copyright notation herein.
    \newline
    Emails: $^\star$\{zhongyuan.zhao, segarra\}@rice.edu, $^\ddag$\{jake.b.perazzone.civ, gunjan.verma.civ, kevin.s.chan.civ, ananthram.swami.civ\}@army.mil
    \newline 
    Source code: \url{https://github.com/zhongyuanzhao/edgeSPBP}
    }
}
\address{$^\star$Rice University, USA \hspace{10mm} $^\ddag$DEVCOM Army Research Laboratory, USA}

\pgfplotsset{compat=1.18}

\begin{document}
\ninept
\maketitle

\begin{abstract}
A significant challenge for computation offloading in wireless multi-hop networks is the complex interactions among traffic flows in the presence of interference.
Existing approaches often ignore these key effects
and/or rely on outdated queueing and channel state information.
To fill these gaps, we reformulate joint offloading and routing as a routing problem on an extended graph with physical and virtual links.
We adopt the state-of-the-art shortest path-biased Backpressure routing algorithm, 
which allows the destination and the route of a job to be dynamically adjusted at every time step based on network-wide long-term information and real-time states of local neighborhoods.
In large networks, our approach achieves smaller 
makespan than existing approaches, such as separated Backpressure offloading, and joint offloading and routing based on linear programming. 
\end{abstract}

\vspace{-0.05in}
\section{Introduction}\label{sec:intro}
\vspace{-0.1in}
Computation offloading (also referred to as mobile edge computing, fog computing, and cloudlets)~\cite{ferrer2019towards} is becoming critical to many network applications, such as Internet-of-Things (IoT), connected vehicles, smart cities, drone swarms, military communications, and disaster relief~\cite{sarkar2013ad,kreutz2014software,cisco2020,kott2016internet}, since it allows rich sensory data from resource-limited mobile devices to be processed in a timely manner at external servers.
In addition, computation offloading can improve the lifespan of battery-powered devices and empower edge intelligence in the age of artificial intelligence (AI), since it is often impractical to equip mobile devices with hardware accelerators due to economic or energy constraints. 
Many of these emerging applications increasingly rely on self-organizing wireless multi-hop networks for connectivity, including xG (device-to-device, wireless backhaul, and non-terrestrial coverage), vehicle-to-everything (V2X), mobile ad-hoc/sensor networks, and machine-to-machine (M2M) communications~\cite{kott2016internet,akyildiz20206g,chen2021massive,noor20226g}.
Therefore, there is a pressing need to develop distributed computation offloading in wireless multi-hop networks,
in which resource allocation does not rely on infrastructure~\cite{tassiulas1992,neely2005dynamic,zhao2023delay,zhao2023enhanced,zhao2024tmlcn,joo2010complexity,joo2012local,Alresaini2016bp,ji2012delay,zhao2022twc,zhao2023graphbased}.

Computation offloading often focuses on reducing the makespan (job initialization to completion time) and/or energy consumption of tasks by deciding where and when each job should be processed as well as how to upload the job description to the server~\cite{ferrer2019towards}.
A popular approach is to use mixed-integer programming (MIP)~\cite{Müller2015computation,funai2019computational,liu2020distributed,feng2021multi,kiamari2022gcnscheduler,li2022novel,dai2022learning}, which seeks to optimally schedule individual jobs, subject to the constraints of their dependency relationships and the residual capacities of servers and communication channels. 
MIP has been applied to scenarios with complex task dependencies~\cite{li2022novel,feng2021multi} and where servers are only one hop away from clients~\cite{li2022novel,dai2022learning,kiamari2022gcnscheduler}.
In multi-hop offloading, however, MIP faces the challenges of 1)~NP-hardness in finding a solution and 2)~timeliness in collecting the residual capacities of servers multiple hops away from the scheduler.
The mean-field linear programming (LP) relaxation~\cite{funai2019computational,liu2020distributed,feng2021multi}, is  more tractable, but often overlooks the key effects of queuing and link scheduling for managing the contentions among traffic flows and transceivers~\cite{zhao2024congestionaware}.

When jobs are recurrent and can be independently processed by a server, stochastic network optimization (SNO) is a better option~\cite{Destounis2016,wang2021latency,Kamran2022deco,lin2020distributed,jiang2023joint,liu2019dynamic,Bi2021Lyapunov}, in which decisions are driven by the queuing state information (QSI) of clients, servers, and relays.
Compared to MIP, SNO can handle  network stochasticity by queuing jobs, while satisfying the mean-field constraints (queue stability) by Lyapunov optimization. 
Originating from routing, SNO is suitable for joint offloading and routing~\cite{Kamran2022deco,lin2020distributed,jiang2023joint,liu2019dynamic,Bi2021Lyapunov}.
However, since the destination of a job is fixed during routing in all the existing schemes~\cite{Müller2015computation,funai2019computational,liu2020distributed,feng2021multi,kiamari2022gcnscheduler,li2022novel,dai2022learning,zhao2024congestionaware,Destounis2016,wang2021latency,Kamran2022deco,lin2020distributed,jiang2023joint,liu2019dynamic,Bi2021Lyapunov}, shifts in network state (including QSI) may render the initial offloading decision suboptimal by the time the job reaches its destination.

To address this challenge, we propose a joint offloading and routing scheme based on shortest path-biased Backpressure (SP-BP) routing~\cite{neely2005dynamic,zhao2023delay,zhao2023enhanced,zhao2024tmlcn}, a special case of SNO. 
Instead of determining the destination of a job upfront, in our approach, the next step of each job --- being queued, processed, or sent to a neighbor --- is dynamically decided by its residing node at every time step, according to network-wide long-term computing and communication capacities and real-time link rates and QSI in its immediate neighborhood.  

\noindent
{\bf Contribution:} The contributions of this paper are threefold:\\
1) For the first time, we incorporate the idea of virtual sink in the field of network flow~\cite{edmonds1972theoretical} to enable joint offloading and routing in multi-hop networks under a unified routing framework.
In particular, we model computation at a server as sending jobs to a virtual sink over a virtual link -- an operation compatible with communications.\\
2) We develop a fully distributed scheme of joint offloading and routing by leveraging the throughput optimality of SP-BP and developing a job scheduler consistent with SP-BP for task computing. \\
3) We demonstrate that our approach achieves a smaller makespan in large networks (of 100 nodes in our simulations) compared to a separated BP offloading and routing scheme and a linear programming (LP)-based mean-field joint offloading and routing scheme.

\begin{figure*}[!t]
    \centering
    \vspace{-0.05in}
    \subfloat[]{
        \includegraphics[height=1.72in]{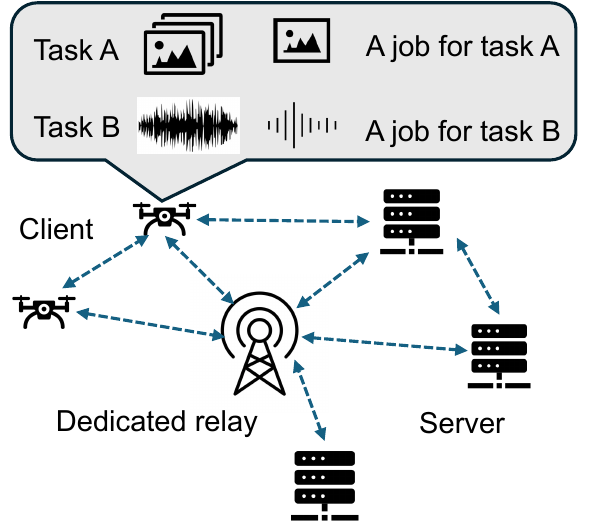}\label{fig:model:wmn}
        \vspace{-0.1in}
    }
    \subfloat[]{
        \includegraphics[height=1.72in]{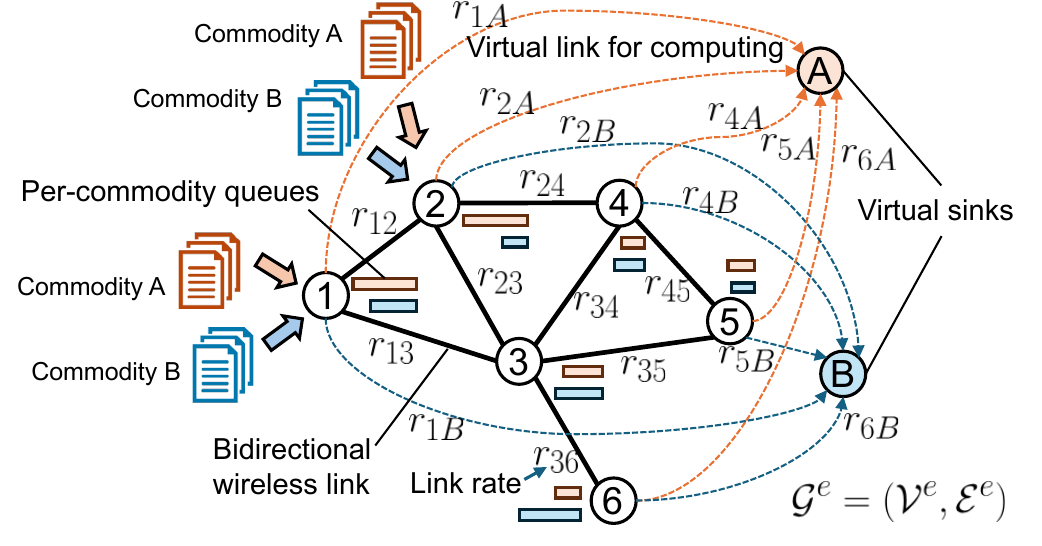}\label{fig:model:system}
        \vspace{-0.1in}
    }
    \subfloat[]{
        \includegraphics[height=1.72in]{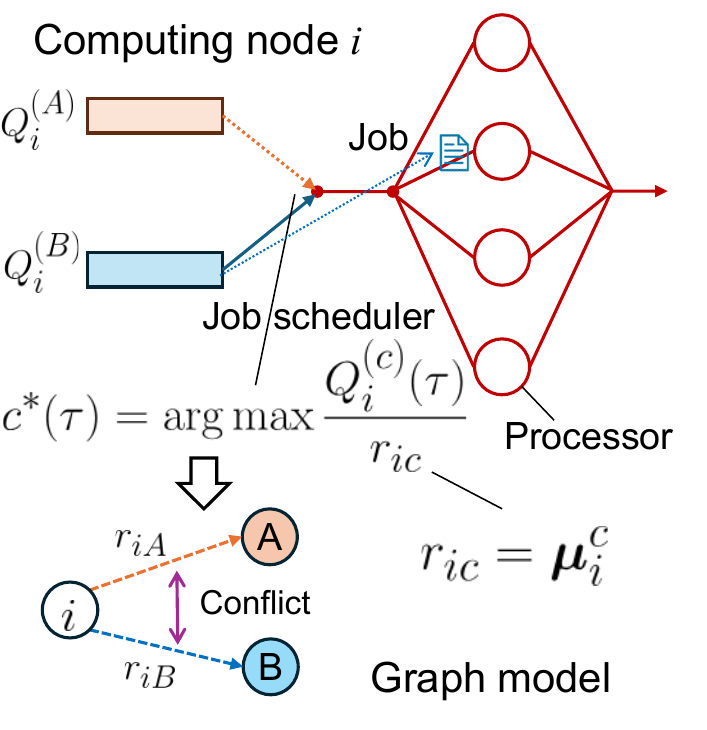}\label{fig:model:compute}
        \vspace{-0.1in}
    }
    \vspace{-0.15in}
    \caption{Graph modeling: (a) Task offloading in a wireless multi-hop network where a client may have two types of tasks.
    (b) Extended graph and per-commodity queues for the example in (a). 
    (c)~Job scheduling for two types of jobs with equal priority at a computing node $i$ with four processors, the computing capacity of a server is modeled as two virtual links, with link rate equal to the corresponding service rate.
    }
    \label{fig:model}
    \vspace{-0.2in}
\end{figure*}

\vspace{-.3mm}
\noindent
{\bf Notation:} 
$ (\cdot)^\top $, $ \odot $, and $ |\cdot| $ represent the transpose operator,      Hadamard (element-wise) product operator, and the cardinality of a set.
$ \mathbbm{1}(\cdot) $ is the indicator function.
$ \mathbb{E}(\cdot) $ stands for expectation.

\section{System Model and Problem Formulation}
\label{sec:problem}

We consider a wireless multi-hop network with three types of devices: mobile clients, servers, and dedicated relay nodes. 
A client is equipped with sensors and limited computing resources, such as IoT sensors, 
wearable devices, or drones.
A server has richer computing resources than clients but may be located multiple hops away from them. 
While all devices can buffer and relay packets, a dedicated relay node cannot initiate or execute any computational tasks. 
An example of such a network with 6 nodes is illustrated in Fig.~\ref{fig:model:wmn}. 

The wireless multi-hop network is modeled as a \emph{connectivity graph} $\ccalG^{n}$ and a \emph{conflict graph}~$\ccalG^{c}$. 
The connectivity graph is a bidirectional graph $\ccalG^{n}=(\ccalV, \ccalE)$, in which $\ccalV$ is a set of nodes representing wireless devices in the network and $\ccalE$ is a set of links in which $e=(i,j)\in\ccalE$ for $i,j\in\ccalV$ indicates that node $i$ can transmit data to $j$ directly.
We assume that $(j,i)\in\ccalE$ if $(i,j)\in\ccalE$, and that 
$\ccalG^{n}$ is connected, i.e., a path always exists between any pair of nodes.
Based on the roles of devices, $\ccalV$ is partitioned into three subsets: $\ccalM$ for clients, $\ccalR$ for dedicated relays, and $\ccalS$ for servers.

The conflict graph, $\ccalG^c=(\ccalE,\ccalH)$, 
describes the conflict relationship between links (here, we re-use the notation $\ccalE$ referring to its elements as vertices) and is defined as follows: each vertex $e\in\ccalE$ corresponds to a link in $\ccalG^{n}$ and each undirected edge $(e_1, e_2)\in\ccalH$ indicates a conflict between links $e_1, e_2\in\ccalE$ in $\ccalG^{n}$.
Two links could be in conflict due to either 1) \emph{interface conflict}, i.e., two links share a wireless device with only one wireless transceiver; or 2) \emph{wireless interference}, i.e., their incident devices are within a certain distance such that their simultaneous transmission will cause the outage probability to exceed a prescribed level.
In this paper, we assume $\mathcal{G}^{c}$ to be known, possibly by each node monitoring the wireless channel~\cite{zhao2022twc}, or through more sophisticated approaches~\cite{yang2016learning}. 

We consider the following task offloading settings. 
A client may generate multiple computational tasks for processing sensor data in real-time, such as object detection in video streams and analysis of audio streams as shown in Fig.~\ref{fig:model:wmn}.
Each task contains a sequence of independent jobs (e.g., object detection for a single video frame, assuming downstream analytics take place somewhere else) 
arriving at its source client at a certain rate, where each job can be individually processed by a single server. 
The tasks in the network are categorized into a set of types, denoted as $\ccalC$, based on the data size and processing algorithm of their jobs. 
The jobs of the same type (e.g., object detection in video streams), regardless of their source nodes, are considered to have identical data sizes (e.g., a video frame) and are processed by the same algorithm (e.g., object detection).
We denote a task of type $c\in\ccalC$ from source $m\in\ccalM$ as $c_m$, its job arrival follows a stationary process with a rate
of $\lambda_{m}^{c}$, and the average makespan of its jobs is $\delta^c_m$. 
We consider a scenario in which all tasks are of equal priority, and the data and description of each job can be enclosed in a single data packet of identical size regardless of its type.
Our approach can be extended to cases where tasks have different priorities, and different types of jobs may contain different numbers of data packets.

We consider a time-slotted medium access control in the wireless network.
Matrix $\bbR\in\reals^{|\ccalE|\times T}$ collects the real-time link rates (capacities)
of all wireless links across $T$ time slots, where $\bbR_{e,t}$ is the instantaneous link rate of $e\in\ccalE$ at time $0 < t\leq T$.
Notice that future link rates, i.e., vector $\bbR_{*,t'}$ where $t'>t$, are not required to be known at time step $t$. 
Vector $\bbr\in\reals^{|\ccalE|}$ collects the long-term average link rates of all wireless links. 
We assume that $\bbR_{ij,t}=\bbR_{ji,t}$ and $\bbr_{ij} = \bbr_{ji}$.
Similarly, vector $\bbmu^c\in\reals^{|\ccalV|}$ describes the service rates of the node set $\ccalV$ for tasks of type $c\in\ccalC$.
Both link rates and service rates are in the unit of number of jobs (packets) per time slot.

In joint offloading and routing, we make two types of decisions: 1) where each job should be processed, e.g., at its source client (locally) or a specific remote server; and 2) through which route the job should be uploaded to the specified server.
Our objective is to minimize the average makespan (communication time and computing time) of all tasks across the network. 
In the mean field, we can model each task $c_m$ as a fluid flow with a rate of $\lambda_m^c$ injected to the network via client $m$, and formulate the problem as optimizing the \emph{static} non-negative flow rate assignments $\{f_m^c(e) | e\in\ccalE, m\in\ccalM, c\in\ccalC\}$, or simply as $ \{f_m^c(e)\} $, 
to minimize the total makespan of all tasks:
\begin{subequations}\label{E:formulation}
	\begin{align}
		\{f_m^c(e)\}^* &= \argmin_{\{f_m^c(e)\}\in\{\reals^+\}} \sum_{m\in\ccalM}\sum_{c\in\ccalC} \delta_m^c \label{E:formulation:obj}\\
		\text{s.t. } 
		  \delta_m^c & = \sum_{e\in\ccalE}f_m^c(e)u(e) + \sum_{v\in\ccalV}g_m^c(v)u^c(v)\;, \label{E:formulation:cost}  \\
		 \lambda^c_{m}\mathbbm{1}\!(m\!=\!v) &+ \!\!\!\!\sum_{i\in\ccalN(v)} \!\!\!\!f_m^c\big((i,v)\big) = g_m^c(v) +\!\!\!\!\sum_{i\in\ccalN(v)} \!\!\!\!f_m^c\big((v,i)\big) , \label{E:formulation:in}\\
		 \lambda^c_{m} &= \sum_{v\in\ccalV}g_m^c(v)\;,\;\forall\; m\in\ccalM, c\in\ccalC\;,  \label{E:formulation:cons} \\
		 \psi(v) &\geq  \sum_{c\in\ccalC, m\in\ccalM} g_m^c(v)h^c(v), \;\forall\; v\in\ccalV   \;, \label{E:formulation:comp} \\ 
		 \psi(e) & \geq \sum_{c\in\ccalC, m\in\ccalM} f_m^c(e), \;\forall\; e\in\ccalE  \;, \label{E:formulation:comm} \\
		 h^c(v), u(e),\; & g_m^c(v), u^c(v)\in\reals^{+},\forall\; e\!\in\!\ccalE, m,v\!\in\!\ccalV, c\!\in\!\ccalC,\label{E:formulation:e}
	\end{align} 
\end{subequations}
where \eqref{E:formulation:in} is for all $ m\in\ccalM, c\in\ccalC, v\in\ccalV$.
In \eqref{E:formulation}, $u(e)$ and $u^c(v)$ are respectively the unit costs of sending flow over link $e$ and processing job flow of type $c$ via node $v$, $\psi(v)$ and $\psi(e)$ are respectively the long-term computing and communication capacities of node $v$ and link $e$ under given scheduling policies, $\ccalN(v)$ is the set of immediate neighbors of node $v$ on $\ccalG^n$, $g_m^c(v)$ is the processing rate of flow $c_m$ on node $v$, and $h^c(v)$ is the unit computing capacity demanded by job flow of type $c$ on $v$.
The constraints in \eqref{E:formulation} are explained as follows:
\eqref{E:formulation:cost} defines the average makespan of task $c_m$ as the sum of communication and computing costs across all the links and nodes under policy $\{f_m^c(e)\}$.
\eqref{E:formulation:in} states the flow conservation law that the inflow of any node must equal  its outflow, and flows must be only on the existing links $e\in\ccalE$. 
\eqref{E:formulation:cons} states that the total processing rate of flow $c_m$ in the network must be equal to its arrival rate, i.e., all jobs must be processed.
\eqref{E:formulation:comp} and \eqref{E:formulation:comm} respectively state that computing on node $v$ and communication over link $e$ are limited by their long-term capacities.
\eqref{E:formulation:e} limits all processing rates, unit capacity demands, and unit costs to be non-negative real values.

Notice that the LP formulation in \eqref{E:formulation} can be solved efficiently only if $u(e), u^c(v), \psi(v), \psi(e)$ for all $e\in\ccalE, v\in\ccalV, c\in\ccalC$ are given as constants.
However, in practice, such unrealistic assumptions often lead to suboptimal solutions, since these unit costs and capacities in wireless networks are non-closed-form  functions of the conflict graph, the decision variables $\{f_m^c(e)\}$, link scheduling scheme, and channel fading~\cite{zhao2024congestionaware}.
Moreover, the need for a centralized solver also limits the scalability and robustness of the LP approach.

\section{Joint Offloading and Routing}\label{sec:solution}
Instead of relying on the formulation in \eqref{E:formulation}, we reformulate the joint offloading and routing as a problem of finding the best route for a job to reach a virtual sink.
A job reaching the virtual sink indicates it has been completed -- by computing on the penultimate node.
A node $v$ processing a job of type $c$ is modeled as the job being transmitted from $v$ to virtual sink $c$ via a virtual link.
This formulation enables us to approach joint offloading and routing using traditional routing algorithms, from which we choose fully-distributed SP-BP routing~\cite{neely2005dynamic,zhao2023delay,zhao2023enhanced,zhao2024tmlcn}.
Our approach is detailed as follows.

\subsection{Graph Modeling and Per-Commodity Queues}
\label{sec:model}
We build an \emph{extended graph} $\ccalG^e=(\ccalV^e, \ccalE^e)$ based on $\ccalG^n=(\ccalV,\ccalE)$ by introducing a virtual sink for each task type $c\in\ccalC$, i.e., $\ccalV^e=\ccalV \cup \ccalC $, and adding a directed virtual link from each computing node (client or server) to each virtual sink, i.e., $\ccalE^e = \ccalE \cup \{(i,c) | i\in\ccalM\cup\ccalS, c\in\ccalC\}$.
The link rate $\bbr^e_{ic} $ of $(i, c)$ is the service rate $ \bbmu_i^c$, defined as the number of jobs of type $c$ that can be processed by node $i$ in a time slot, when node $i$ is free of other types of jobs.

More specifically, jobs of type $c\in\ccalC$, regardless of their sources, are modeled as being commodity $c$ that is destined to virtual sink $c$.
Each node in the network hosts per-commodity queues in a first-in-first-out (FIFO) fashion.
The queue for commodity $c$ hosted on node $i\in\ccalV$ is denoted as $\ccalQ_i^{(c)}$, and its length at the beginning of time slot $t$ as $Q_i^{(c)}(t)$. 
A virtual sink $c\in\ccalC$ only accepts commodity $c$ and $ Q_c^{(c)}(t)=0 $ for any $t$.
The extended graph and the per-commodity queueing system for the exemplary network in Fig.~\ref{fig:model:wmn} are illustrated in Fig.~\ref{fig:model:system}.
The joint offloading and routing is then accomplished by employing SP-BP routing on the extended graph $\ccalG^e$.

\subsection{Job Scheduling at Computing Nodes}
\label{sec:job}
Since job scheduling at a computing node is not synchronized to
the time-slotted system in wireless networks, we design a job scheduler compatible with BP routing and scheduling as follows.
In a multi-processor server $i\in\ccalM\cup\ccalS$, as exemplified in Fig.~\ref{fig:model:compute}, whenever a processor becomes free at time $\tau$, it fetches a job into its memory from the queue of the optimal commodity $ c^*(\tau) $, where:
\vspace{-0.05in}
\begin{equation}\label{E:job}
	c^*(\tau)=\argmax_{c\in\ccalC} \frac{Q_i^{(c)}(\tau)}{\bbmu^c_{i}}.
\vspace{-0.05in}
\end{equation}

This job scheduler emulates the SP-BP in a virtual link as follows: 
commodity $c$ is the only one allowed, thus the optimal one on virtual link $(i, c)$, and the virtual links $\{(i, c) | c\in\ccalC \}$ are only in conflict with each other but not with any wireless links.
Therefore, \eqref{E:job} emulates the MaxWeight scheduler by scheduling virtual link $(i, c)$ with a queue that requires the longest time to be cleared.

\subsection{Shortest Path-biased Backpressure Routing}
\label{sec:bp}
We can now apply the SP-BP algorithm to our extended graph~\cite{zhao2023delay,zhao2023enhanced,zhao2024tmlcn}.
The biased backlog metric of commodity $c$ on node $i$ at $t$ is defined as 
\begin{equation}\label{E:backlog}
	{U}_{i}^{(c)}(t) = Q_{i}^{(c)}(t) + B_{i}^{(c)}  \;,
\end{equation}
where $ 0\leq B_{i}^{(c)}<\infty $ for $i\in\ccalV, c\in\ccalC$, is the queue-agnostic, shortest path distance from $i$ to $c$ on the extended graph $\ccalG^e$ with edge weights $ \sigma_{ij}=\bar{r}r_{max}/\bbr^e_{ij}$ for all $(i,j)\in\ccalE^e $, where $\bar{r}=\mathbbm{E}_{(i,j)\in\ccalE^e}(\bbr^e_{ij})$ and $r_{max}=\max_{(i,j)\in\ccalE^e}(\bbr^e_{ij})$.
$\{B_i^{(j)}|i,j\in\ccalV^e\}$ can be found by the weighted all-pairs-shortest-path algorithm~\cite{bernstein2019distributed}.
BP routing and scheduling consists of 4 steps in each time slot.

\noindent \textbf{Step 1}. The optimal commodity $c_{ij}^{*}(t)$ on each {directed} link (${i,j}$) is selected as the one with the maximal backpressure, i.e., 
\begin{equation}\label{E:commodity}
    c_{ij}^{*}(t) =\argmax_{c\in\ccalV}\{ U_{i}^{(c)}(t) - U_{j}^{(c)}(t) \} \;,
    \vspace{-0.05in}
\end{equation}
where $U_{i}^{(c)}(t)$ is the backlog metric described in \eqref{E:backlog}. 

\noindent \textbf{Step 2}. The maximum backpressure of (${i,j}$) is found as
\begin{equation}\label{E:weight}
    w_{ij}(t) \!=\!\max\!\left\{ U_{i}^{(c_{ij}^*(t))}\!(t) \!-\! U_{j}^{(c_{ij}^*(t))}\!(t), 0 \right\}.
\end{equation}

\noindent \textbf{Step 3}. MaxWeight scheduling \cite{tassiulas1992} finds the schedule $\bbx(t)\in\{0,1\}^{|\ccalE|}$ in order to activate a set of \emph{non-conflicting links} achieving the maximum total utility as follows: 
\begin{equation}\label{E:scheduling}
    \bbx (t) = \argmax_{\tilde{\bbx} (t)\in \ccalX } ~ \tilde{\bbx}(t)^\top  \left[\bbR_{*,t}\odot\tilde{\bbw}(t)\right] \;,
\vspace{-0.1in}
\end{equation}
where vector $\tilde{\bbw}(t)\!=\!\left[ {w}_{ij}\!(t)\mathbbm{1}\!\!\left(\!Q_i^{(c_{ij}^*(t))}\!(t)\!>\!0\!\right) | (i,j)\in\ccalE\right]$, 
$\ccalX$ denotes the set of all non-conflicting configurations, and 
vector $\bbR_{*,t}$ collects the real-time link rates of all links.
MaxWeight scheduling involves solving an NP-hard maximum weighted independent set (MWIS) problem \cite{joo2010complexity} on the conflict graph to find a set of non-conflicting links. 
In practice, \eqref{E:scheduling} can be solved approximately by distributed heuristics, such as local greedy scheduler (LGS)~\cite{joo2012local} and its GCN-based enhancement~\cite{zhao2022twc}.
In our test, LGS is used.

\noindent \textbf{Step 4}. All of the real-time link capacity $\bbR_{ij,t}$ of a scheduled link is allocated to its optimal commodity $c_{ij}^{*}(t)$.
The final transmission and routing variable of commodity $c\in\ccalV$ on link (${i,j}$) is
\begin{equation}\label{E:quota}
    \mu_{ij}^{(c)}(t) \!=\! \begin{cases}
         \bbR_{ij,t}, & \text{if } c=c_{ij}^{*}(t), w_{ij}(t)>0, \bbx_{ij}(t)=1, \\
         0, & \text{otherwise}.
    \end{cases}
\vspace{-0.05in}
\end{equation}
There will be $\min\!\left[\mu_{ij}^{(c_{ij}^*(t))}(t), Q_{i}^{(c_{ij}^*(t))}(t)\right]$ packets being transmitted over link (${i,j}$) at time slot $t$.

Notice that the job scheduling and computing described in Section~\ref{sec:job} operate in continuous time independently from the BP steps in \eqref{E:commodity}-\eqref{E:quota}, sending jobs to virtual sinks in parallel. 
The last-packet problem (a well-established drawback of BP schemes \cite{Alresaini2016bp,ji2012delay}) can also be significantly alleviated since short-lived tasks are merged into only $|\ccalC|$ commodities in our approach instead of $|\ccalS|$ commodities in conventional BP routing (assuming $|\ccalC|\ll |\ccalS|$).

\begin{figure*}[!t]
	\centering
	\vspace{-0.05in}
	\hspace{-2.5mm}
	\subfloat[]{
		\includegraphics[height=1.74in]{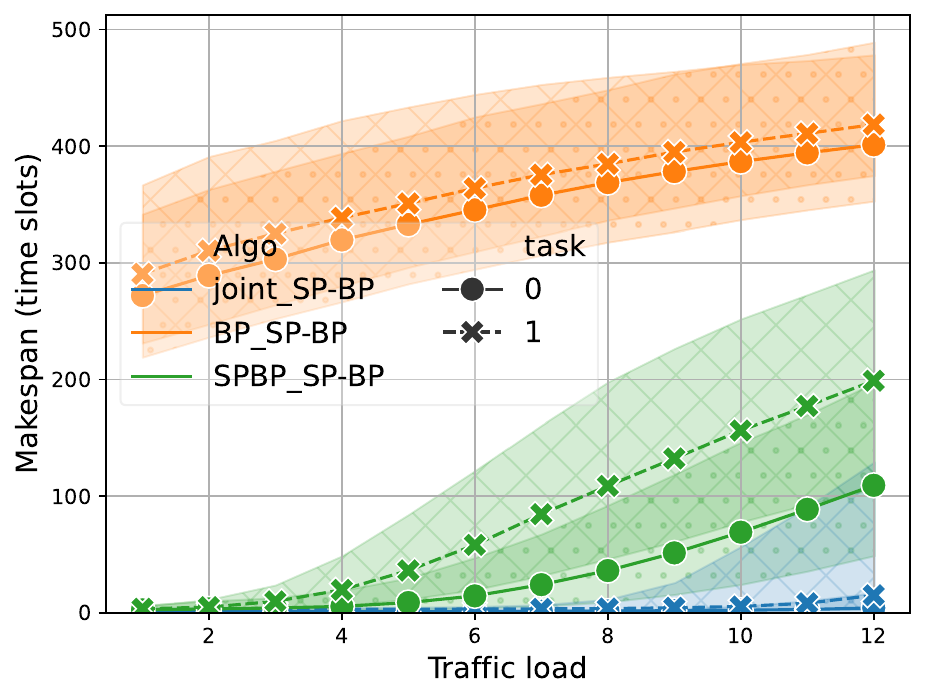}
		\label{fig:results:delay}\vspace{-0.1in}
	}\hspace{-2mm}
	\subfloat[]{
		\includegraphics[height=1.74in]{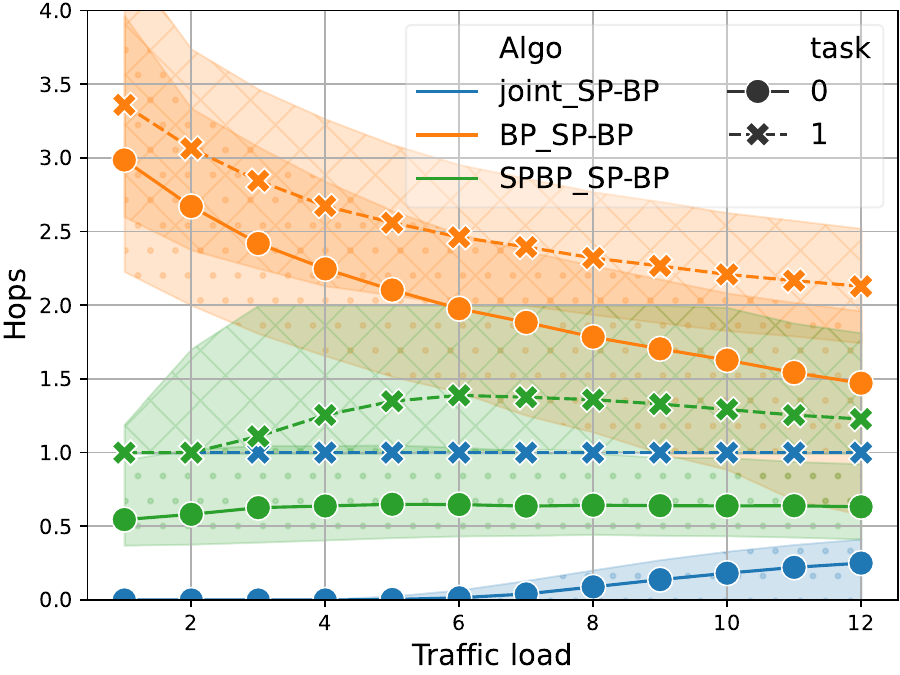}
		\label{fig:results:hops}\vspace{-0.1in}
	}\hspace{-2mm}
	\subfloat[]{
		\includegraphics[height=1.74in]{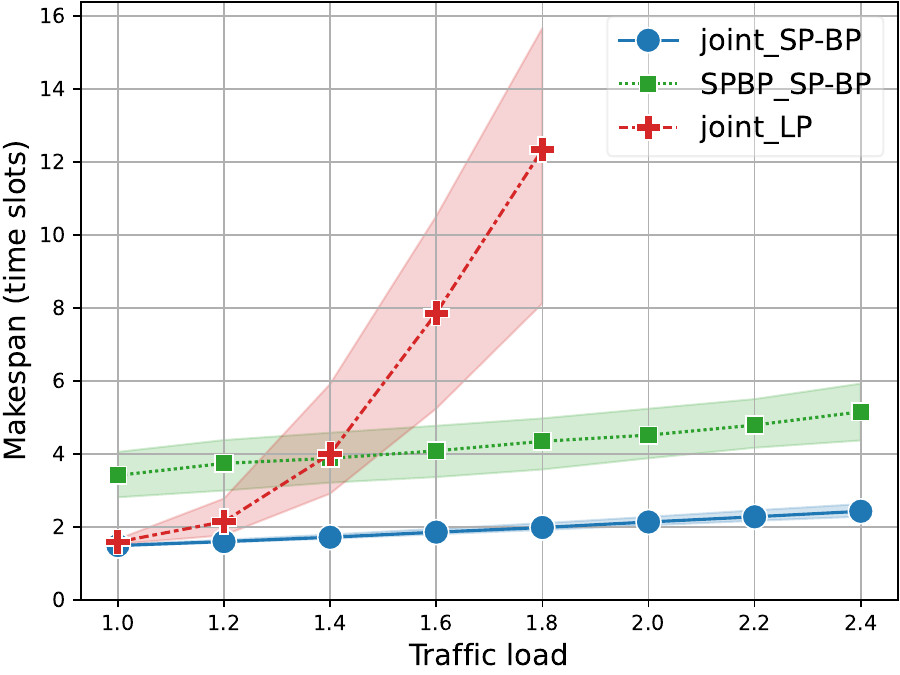}
		\label{fig:results:form}\vspace{-0.1in}
	}
	\vspace{-0.15in}
	\caption{{\small 
			Numerical results on networks of 100 nodes:
			(a)~Median makespan under different schemes as a function of traffic load, and
			(b)~Median number of hops a job traveled before being processed, under the setting of two types of tasks.
			(c)~Median makespan under different formulations as a function of traffic load under a single type of task.
			The bands indicate 25 and 75 percentiles, lines for median. 
		}
	} 
	\label{fig:results}    
	\vspace{-0.1in}
\end{figure*}

\begin{figure}[!t]
	\centering
 \vspace{-0.1in}
	\includegraphics[height=1.15in]{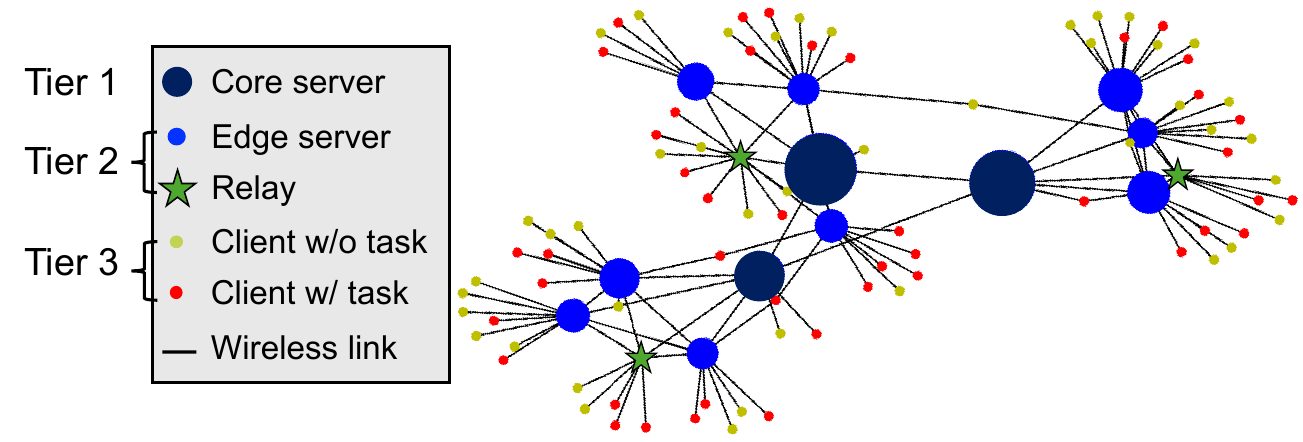}
	\vspace{-0.3in}
	\caption{An exemplary network with $100$ nodes, where node color indicates its role and node size represents computing capacity. 
 }
	\vspace{-0.2in}
	\label{fig:instance}
\end{figure}

\vspace{-0.1in}
\section{Numerical experiments}
\label{sec:results}
\vspace{-0.05in}

Our joint offloading and routing (joint\_SP-BP) and other benchmark schemes are evaluated on 10 randomly generated wireless ad-hoc networks with 100 nodes, as exemplified in Fig.~\ref{fig:instance}. 
Each connectivity graph $ \ccalG^{n} $ has a 3-tier structure: 
first, draw $k\in\{3,4,5,6\}$ with uniform distribution, and create a clique of $k$ core servers (tier-1); 
next, $4k$ tier-2 nodes ($80\%$ edge servers and $20\%$ dedicated relays) in $k$ groups are added, nodes in the $k$th group are all directly connected to the $k$th core server, and connected with each other with a probability of $0.7$;
lastly, each of the $100-5k$ clients (tier-3) is connected to one (or two with a probability of $0.1$) random server(s) (an edge server with a probability of $ 10/(41 k) $ or a core server with a probability of $1/(41k)$). 
The conflict graph $\ccalG^c$ is the line graph of $\ccalG^n$.
The base service rates $\bbmu $ are generated as follows: for servers, $|\ccalS|$ random values are drawn from a Pareto distribution with a shape of 2 and a scale of 8, 
and the $s$th largest element is assigned to the server added to the network in the $s$th order;
for a client $m$, $ \bbmu_m \in \mathbbm{U}(8, 12) $.

On each of the 10 network instances, 10 test instances with a horizon of $T=1000$ are generated as follows.
For each type $c\in\{0, 1\}$, tasks are created on $\lfloor p|\ccalM|\rceil$ random clients, where $ p\in\mathbbm{U}(0.3, 1)$.
The job arrival of task $c_m$ follows a Poisson process with
a rate of $\lambda_m^c=l \tilde{\lambda}_m^{c}$, where $\tilde{\lambda}_m^{c}\in\mathbb{U}(0.5, 1.0) $ and $l$ is the traffic load.
To differentiate the computing resources required by each type of tasks, e.g., $h^c(v)$ in~\eqref{E:formulation}, the service rates are set as $\bbmu^0=0.6\bbmu$, and $\bbmu_v^1=\bbmu_v\mathbbm{1}\!(v\in\ccalS)$ (task type 1 cannot be processed by clients). 
The long-term and real-time link rates $\bbr$ and $\bbR$ are realized the same as \cite{zhao2024tmlcn}.
With a probability of $0.5$, a task $c_m$ is bursty with job arrivals only occurring in $\left[t, t+30\right]$ where $t\in\mathbbm{U}(0, T-200)$. 
Otherwise, jobs for streaming tasks can arrive at any time.

The benchmark schemes are  as follows: 
1) SPBP\_SP-BP: when a new job of task type $c_m$ is created on $m$, its destination is selected as the server with the minimal biased backlog, as follows: 
\vspace{-0.05in}
\begin{equation}\label{E:BPoffloading}
    i^*=\argmin_{i\in\ccalS\cup\{m\}} \left[B_m^{(i)}+B_i^{(c)}+Q_{i}^{(c)}(t)\right]\;,
\vspace{-0.05in}
\end{equation}
and jobs are taken to their destinations by SP-BP routing.
The server backlog $Q_{i}^{(c)}(t)$ is assumed to be instantly available to client $m$ at $t$, and the biases for offloading and routing are the ones in~\eqref{E:backlog}.
2) BP\_SP-BP~\cite{Destounis2016,wang2021latency,Kamran2022deco}: same as SPBP\_SP-BP but the biases in~\eqref{E:BPoffloading} (offloading only) are zeros.
3) joint\_LP: the probabilities for a job of task $c_m$ being sent to a neighbor or processed at its residing node are proportional to the static optimal rates $\{f^c_m(e)\}^*\cup\{g^c_m(v)\}^*$, which are found by solving the LP formulation in \eqref{E:formulation}, assuming that $\psi(v)=\bbmu_v^0$, $\psi(e)=\bbr_{e}/d^c(e)$, $u(v)=1/\psi(v), u(e)=1/\psi(e)$, where $d^c(e)$ is the degree of link $e$ on the conflict graph $\ccalG^c$.

We first compare joint\_SP-BP, BP\_SP-BP, and SPBP\_SP-BP under the setting described before.
In Figs.~\ref{fig:results:delay} and~\ref{fig:results:hops}, we plot the makespan and number of hops a job travels before completion for each of the three schemes. 
BP\_SP-BP performs poorly by only relying on QSI, as evidenced by the large makespan and hop distance of jobs.
SPBP\_SP-BP enhances BP\_SP-BP by considering the distances ($B_m^{(i)}$) to  and computing capacities ($B_i^{(c)}$) of servers in offloading decisions, substantially reducing the offloading distance and makespan.
Our joint\_SP-BP further outperforms SPBP\_SP-BP, confirming the benefit of allowing the route of each job to the virtual sink (e.g., where the job should be processed) to be dynamically adjusted based on accurate network state, rather than deciding it before routing.
The small makespan under joint\_SP-BP shows that it alleviates the last packet problem by merging bursty tasks into only two commodities rather than $|\ccalS|$ in the separated schemes.

Next, we present the makespan under joint\_SP-BP, SPBP\_SP-BP, and joint\_LP in Fig.~\ref{fig:results:form}, for a modified setting with only one task type $\ccalC=\{0\}$, service rates $\bbmu^0=0.125\bbmu$, all tasks are streaming, and everything else the same as the previous test. 
The joint\_LP performs similarly to joint\_SP-BP for low traffic load $l=1.0$, but as $l$ increases, it degrades quickly and underperforms SPBP\_SP-BP, and could not find a valid solution for $l\geq 2.0$. 
This is due to the underlying assumptions of joint\_LP, which deviate further from reality as traffic load increases, and that the static probabilistic policy from joint\_LP cannot react to real-time network congestion as ours does.

\vspace{-0.05in}
\section{Conclusions}
\label{sec:conclusions}
\vspace{-0.05in}

By modeling the computing at a node as sending jobs to virtual sinks over virtual links and leveraging the throughput optimality of Backpressure routing, our fully distributed approach can simultaneously achieve offloading, load balancing, and routing, closing the information gaps in conventional approaches.
Moreover, our approach avoids excessive communication overhead, dependency on a centralized scheduler, and unrealistic assumptions in mean-field LP approaches. 
Potential future directions include incorporating downlink flows, task priorities, and complex task workflows.

\vfill\pagebreak

{\footnotesize
\bibliographystyle{ieeetr}
\bibliography{strings,refs,refs_ol}
}
\end{document}